# User-Centric IT Security

## How to Design Usable Security Mechanisms

Hans-Joachim Hof
Munich IT Security Research Group (MuSe)
Munich University of Applied Sciences,
Lothstraße 64, 80335 Munich, Germany
email: hof@hm.edu

*Abstract*—Nowadays, advanced security mechanisms exist to protect data, systems, and networks. Most of these mechanisms are effective, and security experts can handle them to achieve a sufficient level of security for any given system. However, most of these systems have not been designed with focus on good usability for the average end user. Today, the average end user often struggles with understanding and using security mechanisms. Other security mechanisms are simply annoying for end users. As the overall security of any system is only as strong as the weakest link in this system, bad usability of IT security mechanisms may result in operating errors, resulting in insecure systems. Buying decisions of end users may be affected by the usability of security mechanisms. Hence software providers may decide to better have no security mechanism then one with a bad usability. Usability of IT security mechanisms is one of the most underestimated properties of applications and systems. Even IT security itself is often only an afterthought. Hence, usability of security mechanisms is often the afterthought of an afterthought. Software developers are missing guidelines on how to build security mechanisms with good usability for end users. This paper presents some guidelines that should help software developers to improve end user usability of security-related mechanisms, and analyzes common applications based on these guidelines.

## I. INTRODUCTION

Any improvement of the overall security level of any system requires to improve the security level of all subsystems and available mechanisms as the overall security level of a system is determined by the weakest link in this system [12]. Howe et al. found that current software and approaches for security are not adequate for end users, because these mechanisms are missing ease of use [10]. Arce identifies the end user as weakest link in a company [12]. Hence, improving the usability of security mechanisms helps to improve the overall security level of a system.

Examples of bad usability of security mechanisms are all around. Bad usability of security mechanisms may slow down the adoption of a security system. This happened for example with email encryption. Today, it is very unlikely that an average user uses email encryption. Major problems for average users are key exchange and trust management, both having a very bad usability in common email encryption solutions. Figure 1 shows a completely useless error message during the generation of a key pair for email encryption as one example of bad usability.

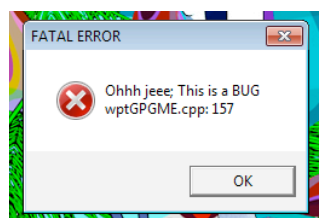

Figure 1. Error message during generation of a key pair for email encryption

The use of email encryption in companies shows that an improved usability may lead to the adoption of the formerly despised technology. In companies, key exchange and trust management are usually not done by the users themselves, but they can rely on central infrastructures such as a central company directory with keys that are trusted by default (all employees). Such a directory ensures average users can use email encryption.

The example of email encryption shows that designing security mechanisms with good usability is worth an effort. For the ordinary software developer, i.e., non security expert, it makes sense not to implement core security mechanisms like encryption algorithms or signature algorithms. Those mechanisms are usually available in security libraries written by security experts and could be easily used by software developers. However, software developers often decide on how security mechanisms are integrated into an application. For example, when implementing an email encryption security solution like GPGMail [11], the software developer decides on the interfaces for setting up trust and importing keys. Both mechanisms are application specific, hence must be implemented by the application developers. Usually, these functionalities are exposed to the users, hence should have a good usability. This paper presents some guidelines that should help software developers to improve end user usability of security-related mechanisms. To underline the importance of the presented guidelines, weaknesses of security mechanisms in common applications regarding usability for end users are shown in an analysis of common applications and security mechanisms on basis of the presented guidelines.

Other important aspects of software security, e.g., secure coding guidelines, testing of security, and threat analysis are out of scope of this paper.

The rest of this paper is structured as follows: Section 2 gives an overview on related work. Section 3 presents guidelines for usable IT security mechanisms. Section 4 analyzes the usability of some common security mechanisms and applications. Section 5 concludes the paper and gives an outlook on future work.

## II. RELATED WORK

Several standards focusing on usability in general exist, e.g., EN ISO 9241 [2]. In EN ISO 9241-11, which is part of EN ISO 9241, requirements for the usability of system are described. These requirements include effectiveness, efficiency and satisfaction. EN ISO 9241-10, another part of EN ISO 9241, lists requirements for usable user dialogs. However, the rules of EN SIO 9241 are very general and not targeted on security mechanisms. The design guidelines presented in this paper interpret the general requirements and rules of EN ISO 9241 and its parts for the special case of security mechanisms.

Other publications like [3][4][5][6][7] focus on the usability of security mechanisms in special applications (e.g., email encryption), or focus on the usability of special security mechanisms (e.g., use of passwords). The guidelines presented in this paper are more general such that they are useful for the design of a wide variety of applications and security mechanisms.

Markotten shows how to integrate user-centred security engineering into different phases of the software development process [1]. However, the emphasize of Markotten's work is more on integration of usability engineering into the software development process than on a design guide.

To summarize, previous works either are not focused on usability of IT security at all or are focused on one special aspect of usable IT security. A set of guidelines for software developers to consider during design of an application is missing. This paper presents some guidelines for software developers to help them improve the usability of security-related functionality.

## III. GUIDELINES FOR GOOD USABILITY OF SECURITY MECHANISMS

The guidelines presented in this section are the result of several years in teaching IT security to beginners (and seeing their difficulties) as well as industrial experience in the design of products requiring IT security mechanisms that are operated by end users. The guidelines reflect our viewpoint on usability of security mechanisms. It is not assumed that those guidelines are complete. It is important to notice that the usability of any system depends on the specific user and his experiences, knowledge and context of use, which includes the task at hand, the equipment at hand, and the physical and social environment of the user. Hence, it is hard to objectively evaluate the usability of a system. However, we hopes that the following set of nine design guidelines coming from the field may be of help for software developers:

**G1 Understandability, open for all users:** As this paper focuses on usability for end users, the average end users should be able to use the security mechanism. Otherwise, the security mechanism is not useful for the intended audience. The average user neither has a special interest in IT security nor understands IT security. It is the responsibility of the software developer to hide as many security mechanisms as possible from the user. For those security mechanisms that are exposed to the end user it is necessary to get security awareness. The process of educating people is easier if suitable metaphors are used. A good metaphor is taken from everyday life of the average user, and is easy to grasp. A good metaphor is simple but powerful in its meaning. Example: an email encryption application should not use the term "encrypted email". It is better to talk about a "secret message for xy" or "email readable only by xy" where xy is the receiver of the message.

Usable security should be available for all users. It should especially not discriminate people. For example, usable security mechanisms should not exclude disabled people that use special tools to access applications (e.g., Braille reader for vision impaired people). Example of compliance with G1: if captchas are used in an application, multiple versions of the captcha should exist. Each version of the captcha should address another sense.

**G2 Empowered users**: Ideally, a usable security mechanism should not be used to restrict the user in what he is doing or what he wants to do. This allows end users to efficiently fulfill their tasks. Efficiency is one of the general usability requirements of EN ISO 9241 [2]. The absence of user restrictions often results in a better acceptance of security by users. The focus of a security mechanism should be on protecting the user. Any security-motivated restriction of the user should be carefully evaluated regarding necessity for system security and adequateness. The user should at least have the impression that he is in control of the system and not the system is controlling him. Security mechanisms should interfere with the usual flow of user activities in the least possible way. Security mechanisms should allow the user to execute activities in any way he wants. Other drivers than protecting the user and the system should not be motivation for restrictions. Especially, users should not be restricted by a security mechanism for the only reason of copyright protection or other business reasons. While such security mechanisms are of great use for businesses, they constantly restrict the user, hence force him to bypass security mechanisms. As users are very imaginative in bypassing unwanted restrictions, it is very likely that a non-security-motivated restriction decreases the security level of a system. The Apple iPhone is a good example: as the phone enforces many restrictions, many user bypass the security mechanisms by using a jailbreak software to revoke those restrictions.

Another important rule is that the user should decide on trust relations. A security mechanism should not enforce trust relations given by a software vendor. The user should always have the possibility to revoke preinstalled trust relations. Trust relations should only be established in advance for the purpose of IT security. For example, having a preinstalled certificate to verify software patches is OK. Establishing trust relations out of business purposes should be avoided. Example of compliance with G2: applications should haven an interface that lists preinstalled certificates. The user should have the possibility to revoke certificates and install custom certificates.

**G3 No jumping through hoops:** Users should only be forced to execute as little tasks as possible that exist only for IT security reasons. Otherwise, users get annoyed and refuse collaboration with IT security mechanisms. The ideal security mechanism does not interfere with user tasks at any time (also see G2). An example on how to not design security mechanisms are captchas: the user is forced to read a nearly unreadable and meaningless combination of letters and numbers and enter it before he can execute the wanted task. Example of compliance with G3: an application that uses a challenge-response mechanism similar to hashcash [9] instead of a captcha to avoid abuse of a service by automated scripts.

**G4 Efficient use of user attention and memorization capability**: Users have problems memorizing data that does not belong to their social background. Hence, they tend to use all kind of optimization to reduce the amount of data they have to remember. This is why users only use approximately 3-4 passwords for all logins where they need passwords. Given the inflationary use of logins in web applications, it is very likely that an average user uses his passwords on multiple sites or for multiple purposes (e.g., for login, for encryption, …). But not only does an average user use the same password more than once, he also selects easy to remember passwords as he is not good in memorizing passwords with a mix of upper and lower case letters, numbers and special characters. Hence, security mechanisms should require the user only to remember little data or no data at all. Example of compliance with G4: An application uses an existing account from another site for login, e.g., by using OpenID [8]. The user can use an existing account, hence does not have to remember another password.

Security mechanisms should only require as little interaction with the user as possible. The security mechanism should only requests the attention of the user if it is absolutely necessary. Interaction with the user should be done in the most minimalistic way. See also G1 for user interaction. Example of compliance with G4: an email encryption application that does not ask a user for each mail if he wants to encrypt the mail or not. Instead, the email application offers a configuration option to always encrypt mails. Additionally, the email composition window clearly states the current protection status and offers a possibility to override the preferences.

**G5 Only informed decisions:** A user only feels secure and cooperates with a system if the system does not ask too much of him. Hence, users should only have to make decisions they can decide on. If there is an important security decision to take, it must be ensured that the user has the capability to make this decision. This means that the user has enough information about the situation that requires him to make a decision, and it must be ensured that the average user is capable to make an informed decision on this issue. If it is not clear if the user can decide on an issue, the decision should be avoided. G5 is hard to achieve and requires a careful examination during the design of an application.

**G6 Security as default:** Good usability requires efficiency. Hence, the user should not have to configure security when he first starts an application. Software for end users should always come preconfigured such that the software is reasonable secure and usable. All security mechanisms of a system should be delivered to the end user with a configuration that offers adequate security for the end users. The configuration effort must be minimized for users. This requires an analysis of the security requirements of average users during software development prior to the deployment of the software to find the adequate security level for most users. Example of compliance with G6: a home wifi access point comes preconfigured with a random WiFi password.

**G7 Fearless System:** The security system should support a positive attitude of the user towards the security system. A user with a positive attitude towards security mechanisms is cooperative and more likely to not feel interrupted by security mechanisms. Hence, security mechanisms should protect the overall system in a way that the user neither has fear when the system is in a secure state nor feels secure when the system is not in a secure state. The security state of the system should be visible at all times. A security mechanism should be consistent in its communication with its user. A security mechanism should not use fear to force users to obey security policies or get a wanted reaction. G7 is hard to achieve and requires a careful examination during the design of an application.

**G8 Security guidance, educating reaction on user errors:** Users tend to make mistakes, especially in respect to IT security. It is important that the security system hinders the user to make mistakes. However, as blocked operations can be very frustrating for users, the response of the security system must provide information why a given operation was blocked and should also offer a solution on how the user could proceed. The solution must be adapted on the situation and should keep the overall security of the system in mind. A security system should guide the user in the usage of security mechanisms. Errors should be prevented and there should be ways to "heal" errors. Example of compliance with G8: when an email encryption application fails to encrypt an email because of a missing public key of the recipient, the error message should explain how to import certificates from and how to verify certificates by comparing fingerprints of keys. To "heal" the error, the email encryption application offers to send the mail as password-protected PDF and instruct the user to call the recipient and tell him the password for the PDF.

**G9 Consistency:** Consistency allows users to efficiently fulfill their tasks. Security mechanisms should fit into both the application and the system context where they are used. Security mechanisms should have the look and feel the user is used to. G9 is hard to achieve and requires a careful examination during the design of an application.

## IV. ANALYSIS OF THE USABILITY OF COMMON SECURITY MECHANISMS AND APPLICATIONS

In this section common applications and security mechanisms are analyzed on basis of the guidelines given in Section III. The analysis identifies room for improvement in these applications and security mechanisms. It also shows some good examples for certain aspects of security usability.

### A. E-Mail Encryption using GPGMail

The encryption process itself is fairly easy, usually requiring one click to enable email encryption. However, key and trust management requires significant effort. For a secure exchange of public keys, the user has to get the public key itself (e.g., from a key server or the homepage of the receiver of a message) and verify the authenticity of the key. Certificates may be in use. The authentication requires the use of another channel to communicate with the key owner (e.g., telephone or in person) and to read a number to the owner that is meaningless for the user. There is no guidance for this process. Then, the user has to change the trust of the exchanged public key. It gets more complicated when using a web of trust for trust management: for the web of trust to work, the user must decide on how trustworthy a person is to verify public keys/certificates in addition to managing direct trust into keys. The distinction between those different types of trust is very hard to understand for average users.

This application is compliant with the following guidelines:
- G2 (user decides on trust relations)
- G4 (minimal interaction)
- G7 (does not frighten user)
- G9 (usually good integration, depends on system, mail client)

This application is not compliant with the following guidelines:
- G1 (hard to understand trust management and process of key verification)
- G3 (complicated trust management)
- G5 (hard to understand trust management and process of key verification)
- G6 (not set to "encrypt all" by default)
- G8 (not much guidance with trust management)

### B. Forced Updates

Keeping a system up-to-date requires a timely use of provided security patches. However, many users are quite lax in applying security patches. Hence, nowadays more and more software providers let not the users decide on when to patch a system but automatically apply security patches as soon as available. While this relieves the user from applying patches, it does not take into consideration the situation of the user at the moment of a forced update. The update process may require downloading a large amount of date. This is a problem when the user is temporary on a low-bandwidth connection. The update process may change security or trust relevant configuration of the application, e.g., by revoking certificates or adding new certificates that are considered trustworthy by the software provider. Often, forced updates cannot be stopped by the user, hence hinder the user.

This security mechanism is compliant with the following guidelines:
- G1 (easy to understand)
- G5 (no user decisions involved)
- G6 (keeps system up-to-date)
- G7 (does not frighten user)
- G8 (no user action necessary (or possible))
- G9 (well integrated)

This security mechanism is not compliant with the following guidelines:
- G2 (user can not decide to not apply a patch, user can not decide on time to apply patch (e.g., do not patch presentation application before presentation on CENTRIC 2012))
- G3 (in some cases user has to wait until patch was applied)
- G4 (full attention of the user when waiting for process to finish)

### C. Captchas

A captcha is a security mechanism avoiding that services are used by automated scripts. In theory, a captcha should be designed in a way that only humans can solve the given problem. Common captcha design requires users to read a distorted and meaningless combination of letters and numbers and enter it before he can use the service. **Figure 2** shows a captcha that is even worse from a usability point of view. Another side effect of the use of captchas is that captchas usually discriminate against disabled people (e.g., vision impaired people).

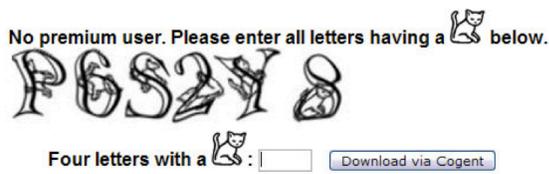
Figure 2. Complicated captcha

This security mechanism is compliant with the following guidelines:
- G5 (no user decision needed)
- G6 (always used)
- G7 (does not frighten user)
- G8 (gives instructions on how to use it)

This security mechanism is not compliant with the following guidelines:
- G1 (discriminates against disabled people)
- G2 (does not allow users to use automation tools)
- G3 (additional task without value for the user)
- G4 (unnecessary user interaction)
- G9 (many different kinds of captchas are in use)

D. *HTTPS Certificate Validation in Common Browsers*

HTTPS allows for confidential and integrity protected communication on the web. For example, HTTPS is used for online banking or shopping. Nowadays HTTPS is widely used on the web. However, for a secure communication it is necessary to avoid man-in-the-middle attacks. To do so, certificates are used to authenticate the web site that one communicates with. As it is not practicable to install a certificate for each and every web site one visits, most common browsers come with preinstalled certificates of so-called Certificate Authorities (CAs). A browser accepts all certificates that have been signed by such a CA. For example, Mozilla Firefox version 14.0.1 comes with over 70 preinstalled CA certificates. The browser software developer decides on the trustworthiness of a CA (and hence on the trustworthiness of web sites), not the end user.

**Figure 3** shows a typical error message of Firefox when encountering a certificate signed by an unknown CA. The text of this error message is too complicated for average users. Above this, average users are not capable of deciding on the validity of unknown certificate anyway. As this error often occurs, the users get used to it and usually just add a security exception to the system to access the web site, bypassing the security mechanism. Adding a security exception involves multiple steps (see **Figure 4** for a screenshot of the second page of the error message when clicking on "Add Exception".

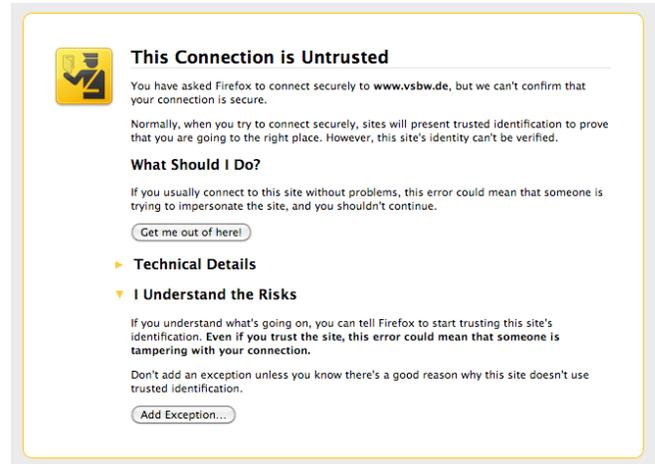
Figure 3. Typical error message of Firefox when encountering an unknown certificate

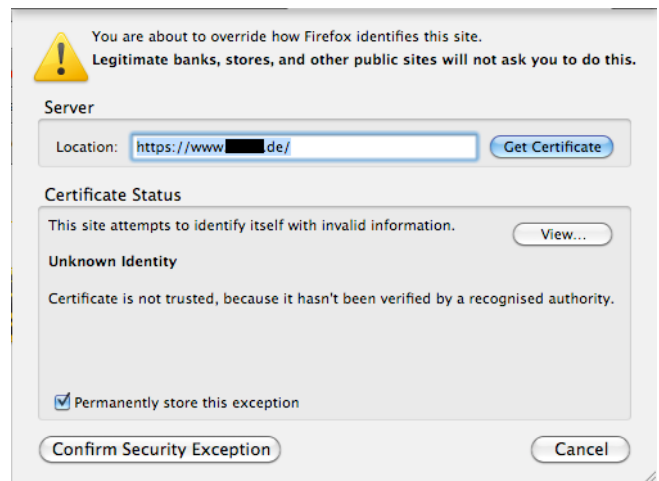
Figure 4. Second dialogue page if user clicked "Add Exception"

This security mechanism is compliant with the following guidelines:
- G6 (large number of preinstalled CAs for secure communication)
- G8 (guidance is given, however the texts used are not suited for average users)

This security mechanism is not compliant with the following guidelines:
- G1 (hard to understand error message given when browser encounters an unknown certificate / a certificate from an unknown CA)
- G2 (many preinstalled CA certificates, the user does not initially decide on trust relations. However, expert users can change the trust settings)
- G3 (annoying additional tasks when unknown certificate / a certificate from an unknown CA is encountered)

- G4 (error unknown certificate happens often, hence most users simply ignore the message and add a security exception)
- G5 (no informed decision possible)
- G7 (error message unknown certificate implies an ongoing attack)
- G9 (look and feel is not consistent with the rest of the browser)

## V. Conclusion and Future Work

This paper presented guidelines for software developers to improve the usability of security-related mechanisms. The analysis of security mechanisms in common applications showed weaknesses in the usability of security-related mechanisms as well as good examples of security usability.

Future work will include the design of usable security mechanisms for common problems, e.g., certificate handling and trust management as well as a user satisfaction study on the effectiveness of the guidelines. The applicability of the guidelines will be checked with software developers that are no security experts. The guidelines presented in this paper are focused on usability for the end user. Future design guides will also focus on better usability for other groups, e.g., system administrators, testers, and developers.